\documentclass[journal,twoside,web]{ieeecolor}
\usepackage{generic}
\usepackage{cite} 
\usepackage{amsmath,amssymb,amsfonts}
\usepackage{algorithmic}
\usepackage{subfig}
\usepackage{graphicx}
\usepackage{booktabs}
\usepackage{textcomp}
\usepackage[pagebackref,breaklinks,colorlinks]{hyperref}
\def\BibTeX{{\rm B\kern-.05em{\sc i\kern-.025em b}\kern-.08em
    T\kern-.1667em\lower.7ex\hbox{E}\kern-.125emX}}
\begin{document}
\title{Deep Reinforcement Learning for Beam Angle Optimization of Intensity-Modulated Radiation Therapy}

\author{Peng Bao, Gong Wang, Ruijie Yang, Bin Dong 
\thanks{This work was supported by the National Natural Science Foundation of China under Grant 12090022, 11831002, 71704023, Beijing Natural Science Foundation under Grant Z180001, the National Key Research and Development Program under Grant 2020YFE0202500, and Beijing Municipal Commission of Science and Technology Collaborative Innovation Project under Grant Z221100003522028. (Corresponding authors: Ruijie Yang, Bin Dong.)}
\thanks{Peng Bao is with the Center for Data Science, Peking University, and the Department of Radiation Oncology, Peking University Third Hospital, Beijing, China. (e-mail: pengbao7598@gmail.com)}
\thanks{Gong Wang, Ruijie Yang is with the Department of Radiation Oncology, Peking University Third Hospital, Beijing, China. (e-mail: wanggong0924@gmail.com, ruijyang@yahoo.com)}
\thanks{Bin Dong is with Beijing International Center for Mathematical
Research \& Center for Machine Learning Research \& National Biomedical Imaging Center, Peking University, Beijing, China. (e-mail: dongbin@math.pku.edu.cn).}
}

\maketitle

\begin{abstract}

\textit{Objective:} Intensity-modulated radiation therapy (IMRT) beam angle optimization (BAO) is a challenging combinatorial optimization problem that is NP-hard. In this study, we aim to develop a personalized BAO algorithm for IMRT that improves the quality of the final treatment.
\textit{Methods:} To improve the quality of IMRT treatment planning, we propose a deep reinforcement learning (DRL)-based approach for IMRT BAO. We consider the task as a sequential decision-making problem and formulate it as a Markov Decision Process. To facilitate the training process, a 3D-Unet is designed to predict the dose distribution for the different number of beam angles, ranging from 1 to 9, to simulate the IMRT environment. By leveraging the simulation model, double deep-Q network (DDQN) and proximal policy optimization (PPO) are used to train agents to select the personalized beam angle sequentially within a few seconds.
\textit{Results:} The treatment plans with beam angles selected by DRL outperform those with clinically used evenly distributed beam angles. For DDQN, the overall average improvement of the CIs is 0.027, 0.032, and 0.03 for 5, 7, and 9 beam angles respectively. For PPO, the overall average improvement of CIs is 0.045, 0.051, and 0.025 for 5, 7, and 9 beam angles respectively.
\textit{Conclusion:} The proposed DRL-based beam angle selection strategy can generate personalized beam angles within a few seconds, and the resulting treatment plan is superior to that obtained using evenly distributed angles.
\textit{Significance:} A fast and automated personalized beam angle selection approach is been proposed for IMRT BAO.
\end{abstract}



\section{Introduction}
\label{sec:introduction}
\IEEEPARstart{C}{ancer} is the second leading cause of death worldwide. According to estimates from World Health Organization (WHO), an estimated 19.3 million new cancer cases and almost 10 million cancer deaths occurred in 2020 \cite{sung2021global}. Therefore, it is significant to improve cancer treatment. Radiotherapy is a primary way to treat cancer and is used in more than half of cancer treatments. It uses high doses of radiation to kill cancerous cells. In this paper, we focus on Intensity-modulated radiation therapy (IMRT), a primary technique for radiotherapy. 

IMRT treatment planning typically involves three steps: 1) tuning the hyper-parameters for each patient, including optimization objectives and beam angles, etc., 2) optimizing the fluence map corresponding to each beam angle, known as \emph{fluence map optimization} (FMO), 3) sequencing the multileaf collimator to generate the deliverable treatment plan. The selection of hyper-parameters is crucial, as it directly affects the quality of the treatment plan and ultimately the effectiveness of the treatment \cite{pugachev2001role}. However, finding the optimal combinations of hyper-parameters for IMRT treatment planning is a challenging and time-consuming task. In current clinical practice, these hyper-parameters are usually obtained by a lot of trial-and-error by the experienced medical physicist, which is time-consuming and makes the quality of the IMRT treatment plan limited by the expertise of the medical physicist. Therefore, there is a significant need to develop an automatic approach to determine the hyper-parameters to generate high-quality treatment plans \cite{wang2020review}.

Numerous studies have been conducted to address the issue of determining the optimal hyper-parameters for IMRT treatment planning. For optimization objectives, one of the most used approaches was to optimize the fluence map with a predefined set of optimization objectives \cite{wang2017development}. Additionally, heuristic and statistical approaches have also been utilized to find better optimization objectives \cite{lee2013predicting, boutilier2015models, wahl2016physically}. More recently, deep reinforcement learning (DRL) has been utilized to tune the optimization objectives based on the current state to mimic the decision-making process of a human planner \cite{shen2019intelligent, shen2020operating}. Specifically, the DRL agent takes the dose-to-volume histogram (DVH) of a treatment plan as the state and determines the action of a specific parameter to improve the plan quality. However, this approach requires a separate neural network for each parameter of optimization objectives, which can result in a large number of networks and is impractical for clinical use. To address this issue, the hierarchical DRL has been introduced to break down the decision-making process into three parts, including the structure, the parameter of the selected structure, and the action to that parameter \cite{shen2021hierarchical}.

In addition to choosing the optimization objectives, beam angle optimization (BAO) has also been studied extensively \cite{hou2003beam, schreibmann2004multiobjective, li2004automatic, dias2014genetic, bortfeld1993optimization, lu1997optimized, djajaputra2003algorithm, woudstra2000constrained, gaede2004algorithm, schreibmann2005dose, woudstra2005automated, vaitheeswaran2010algorithm, dias2014genetic, sadeghnejad2020fast, sadeghnejad2021reinforcement, jia2011beam, yang2006beam, wang2021investigation}. Given the optimization objectives, the BAO problem can be viewed as a combinatorial optimization problem that aims to find the optimal set of beam angles of a given size among all possible angles for the best treatment plan quality. One way to obtain the optimal set of beam angles is to enumerate all possible beam angle combinations, but it is impossible to apply this approach in practice because of the large number of possible combinations. For example, selecting the beam set with 5 beam angles among 360 candidate beam angles results in more than 4.9$e$10 possible combinations, and it will take several minutes to solve the FMO problem to evaluate the plan quality for a given combination of beam angles. As a result, heuristic approaches such as genetic algorithms \cite{hou2003beam, schreibmann2004multiobjective, li2004automatic, dias2014genetic} and simulated annealing algorithms \cite{bortfeld1993optimization,lu1997optimized, djajaputra2003algorithm} have been applied to search for the optimal combination. However, most of these methods usually take a long time because of the huge search space and the global optimality of these methods cannot be guaranteed. Besides, these methods always solve the FMO problem to evaluate corresponding treatment plans which is extremely time-consuming. Instead of solving the FMO problem, some approaches use predetermined criteria to rank the possible beam angles and select the beam angles with the highest scores \cite{woudstra2000constrained, gaede2004algorithm, schreibmann2005dose, woudstra2005automated, vaitheeswaran2010algorithm}. Although these methods are computationally efficient, the optimality of the selected beam angles cannot be guaranteed since the predetermined criteria are based on the single beam and do not adequately consider the interplay between multiple beams. More recently, deep learning approaches have been applied to address the BAO problem. For instance, some works use a deep neural network (DNN) as a surrogate model to quickly estimate the true objective function value of the FMO problem \cite{dias2014genetic, sadeghnejad2020fast}. Moreover, \cite{sadeghnejad2021reinforcement} uses a pre-trained DNN to estimate the beams' fitness values to guide the traversal of branches of the Monte Carlo decision tree to select the beam angle. These methods can reduce the time required to evaluate the treatment plan quality and provide more accurate estimations.

Inspired by the successful application of the DRL algorithm in solving the combinatorial optimization problem \cite{mazyavkina2021reinforcement}, such as the travelling salesman problem (TSP) \cite{bello2017neural, khalil2017learning}, vehicle routing problem (VRP) \cite{nazari2018reinforcement, kool2018attention}, minimum vertex cover problem \cite{khalil2017learning}, scanning strategy \cite{pineda2020active, shen2020learning}, etc., in this work, we propose to use the DRL algorithms to solve the BAO problem. The proposed method can select the beam angles sequentially in just a few seconds based on the patient's anatomy information, the selected beam angles, and the DVH of the current treatment plan. Specifically, we first formulate the BAO problem as a sequential decision problem and design the corresponding Markov Decision Process (MDP). Then we use DRL algorithms, including double deep-Q network (DDQN) \cite{van2016deep} and proximal policy optimization (PPO) \cite{schulman2017proximal}, to train the RL agent to select the beam angles sequentially. Additionally, to obtain the next state and accelerate training, we design a neural network that can predict the dose distribution for different beam settings to simulate the IMRT environment. 

The structure of this paper is organized as follows: in Section \ref{sec:background}, we provide a brief review on IMRT treatment planning and reinforcement learning. Then, in Section \ref{sec:method}, we elaborate on the MDP for IMRT BAO and IMRT environment simulation model. Next, we present the experimental setup and representative results in Section \ref{sec:experiments}. Finally, in Section \ref{sec:conclusion}, we include the conclusion.

\section{Background}
\label{sec:background}

\subsection{IMRT Treatment Planning}
IMRT, as an advanced technique of radiation therapy, uses multiple beams with modulated beam intensity to deliver a more precise radiation dose to the tumor. Given the optimization objectives, the primary goal of IMRT treatment planning is to determine the optimal beam angle set and the corresponding fluence maps to achieve the best possible treatment outcome.

\begin{figure}[t]
    \centering
    \includegraphics[width=2in]{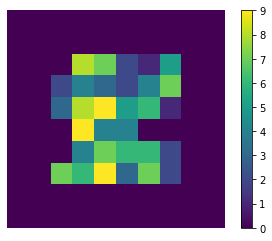}
    \caption{Illustration of fluence map ($10\times 10$). Each square denotes a beamlet.}
    \label{fig:fluence}
\end{figure}

In IMRT, the total accumulated dose of the patient's $i$-th voxel can be calculated as follows:
\begin{equation}
    d_i = \sum_{b\in\mathcal{B}}\sum_{j_b}D_{ij_b}x_{j_b},
\end{equation}
where $i=\{1,2,\dots,I\}$ denotes the index of the voxels, $b\in\mathcal{B}$ represents the beam angle and $\mathcal{B}$ is the set of all candidate beams, $j_b=\{1,2,\dots,J_b\}$ denotes the index of the beamlet at beam angle $b$. $D$ is the dose influence matrix that takes hours to compute, and $D_{ij_b}$ denotes the dose absorbed by $i$-th voxel of the patient when per unit intensity of radiation is delivered through $j_b$-th beamlet at angle $b$. $x$ is the intensity map as illustrated in Figure \ref{fig:fluence} and $x_{j_b}$ is the intensity of $j_b$-th beamlet at angle $b$. The IMRT optimization problem can be formulated as follows:
\begin{equation}
    \begin{aligned}
        \underset{x, \theta}{\min}\ \  & f(d) \\
        {\rm s.t.}\ \  & d_i = \sum_{b\in\mathcal{B}}\sum_{j_b} \theta_b D_{ij_b}x_{j_b} \quad\quad \forall i=1, 2,\cdots, I, \\
        & x \geq 0,\ \  \theta_b\in\{0, 1\},\ \ \sum_{b\in\mathcal{B}}\theta_b\leq B, \\
    \end{aligned}
    \label{formula:FMO_BAO}
\end{equation}
where $\theta_b\in\{0,1\}$ indicates whether the $b$-th angle is selected or not, $B$ is the max number of beam angles to be selected. $f(\cdot)$ denotes a function that quantifies the quality of the treatment plan. There are lots of ways to define $f(\cdot)$, such as linear model\cite{romeijn2003novel}, mixed integer model \cite{lee2003integer}, and multi-objective model \cite{craft2006approximating}. In this paper, we consider the following nonlinear function:
\begin{equation}
    f(d) = \sum_{o\in \mathcal{O}}\left(\alpha_o\left\|d_{i\in \mathcal{V}_o} - \lambda_o p\right\|_+^2+\beta_o \left\|d_{i\in \mathcal{V}_o} - \lambda_o p\right\|_-^2\right),
    \label{formula:f}
\end{equation}
where $\mathcal{O}$ denotes the set of planning target volume (PTV) and organs-at-risk (OARs), $\alpha_o, \beta_o,\lambda_o$ denotes the optimization objective of the organ $o$ which are decided manually by the medical physicist. $\mathcal{V}_o$ denotes the set of the voxels of the organ $o$, and $p$ is the prescription dose. $\left\|\cdot\right\|_+$ and $\left\|\cdot\right\|_-$ are $l_2$-norms that only compute the positive and negative elements respectively.

In current clinical practice, the beam angles are given by the human planner. Given a set of beam angles, the FMO problem can be formulated as follows:
\begin{equation}
    \begin{aligned}
        \underset{x}{\min}\ \  & f(d) \\
        {\rm s.t.}\ \  & d_i = \sum_{b\in\mathcal{B}}\sum_j D_{ij}^bx_{j}^b \ \ {\rm for\ all\ }i=1, 2,\cdots, I, \\
        & x \geq 0. \\
    \end{aligned}
    \label{formula:FMO}
\end{equation}
Since it is time-consuming to solve (\ref{formula:FMO}), in this paper, we propose to use a neural network to directly approximate the corresponding dose distribution instead of solving the FMO problem. As for beam angle optimization, the challenging combinatorial problem, we consider using the DRL algorithm to solve it based on the patient's anatomy information and the current treatment plan.

\subsection{Reinforcement Learning}
A sequential decision problem can be described as a Markov Decision Process (MDP), which is a tuple $\left(\mathcal{S,A}, r, \mathbb{P}, \gamma \right)$ \cite{sutton2018reinforcement}. Here, $\mathcal{S}$ is the state space, $\mathcal{A}$ is the action space, $r:\mathcal{S\times A}\to \mathcal{R}$ is the reward function, $\mathbb{P}:\mathcal{S\times A\times S}\to[0,1]$ is the transition probability of the environment, and $\gamma$ is the discount factor. A policy $\pi:\mathcal{S\times A}\to[0,1]$ is a probability distribution of actions $\mathcal{A}$ over states $\mathcal{S}$. Define a trajectory $\tau=(s_0,a_0,r_0,s_1,a_1,r_1,\dots)$ is the sequence of states, actions, and rewards, the discounted accumulated reward of a trajectory $\tau$ is defined as $G(\tau)=\sum_{t=0}^\infty\gamma^tr_t$.

Given a policy $\pi$, the state-value function $V^\pi(s)$ is defined as the expected return starting from state $s$ and following policy $\pi$:
\begin{equation}
    V^\pi(s)=\mathbb{E}_\tau[G(\tau)|s,\pi].
\end{equation}
Similar to the state-value function $V^\pi(s)$, the state-action-value function $Q^\pi(s,a)$ is defined as the expected return starting from state $s$, taking action $a$, and then following policy $\pi$:
\begin{equation}
    Q^\pi(s,a)=\mathbb{E}_\tau[G(\tau)|s,a,\pi].
\end{equation}
Given an MDP, the RL agent tries to find the optimal policy $\pi^*$ that maximizes the discounted accumulated reward. The algorithms to find the optimal policy $\pi^*$ can be divided into two categories: value-based methods and policy gradient methods. 

Value-based methods, such as Deep Q-Network (DQN) \cite{mnih2015human} use a parameterized neural network $Q(s,a;\psi)$ to approximate the optimal state-action-value function $Q^*(s,a)$ and optimize the neural network by minimizing the following loss function:
\begin{equation}
    \mathcal{L}(\psi) = \mathbb{E}_{s,a,r,s'}\left[\left(Q(s,a;\psi) - (r+\gamma \max_{a'}Q(s',a';\psi'))\right)^2\right],
\end{equation}
where $\psi'$ are the parameters used to compute the target and are copied periodically from $\psi$. However, DQN suffers from an overestimation problem. Double DQN (DDQN) \cite{van2016deep} combines double q-learning \cite{hasselt2010double} and DQN to alleviate this problem, and has the following loss function:
\begin{equation}
    \begin{aligned}
    \mathcal{L}(\psi) = &\mathbb{E}_{s,a,r,s'}[(Q(s,a;\psi) \\
    &- (r+\gamma \max_{a'}Q(s',\underset{a'}{\rm arg\ max}\ Q(s',a';\psi);\psi')))^2].
    \end{aligned}
\end{equation}
With the approximated optimal Q function, the policy can be obtained by $\pi(s)=\underset{a}{\rm arg\ max}\ Q(s,a;\psi)$.

In contrast to value-based methods, policy gradient methods directly optimize a parameterized policy $\pi(a|s;\psi)$ by gradient ascent on $\mathbb{E}_\tau[G(\tau)]$. We use ${\pi_\psi}$ to denotes $\pi(a|s;\psi)$ for simplicity. Proximal policy optimization (PPO) \cite{schulman2017proximal} is one of the state-of-the-art policy gradient methods that updates the policy in a proximal region to address the high variance problem in the vanilla policy gradient. Specifically, defining the advantage function as $A^{\pi_\psi}(s,a)=Q^{\pi_\psi}(s,a)-V^{\pi_\psi}(s)$. Given an old policy $\pi_{\psi_{old}}$, the policy ratio between the new policy and old policy is $q(\psi)=\frac{\pi_\psi}{\pi_{\psi_{old}}}$. PPO optimizes $\pi_\psi$ by maximizing the following objective:
\begin{equation}
    \begin{aligned}
        & \mathcal{L}(\psi) = \\
        &\mathbb{E}_{s,a\sim \pi_{\psi_{old}}} \left[ \min(q(\psi)A^{\pi_{\psi_{old}}}, {\rm clip}(q(\psi), 1-\epsilon, 1+\epsilon)A^{\pi_{\psi_{old}}}) \right],
    \end{aligned}
\end{equation}
where ${\rm clip}(q(\psi), 1-\epsilon, 1+\epsilon)$ restricts the policy ratio $q(\psi)$ to $[1-\epsilon, 1+\epsilon]$.

\section{Method}
\label{sec:method}
\begin{figure*}
    \centering
    \includegraphics[width=6in]{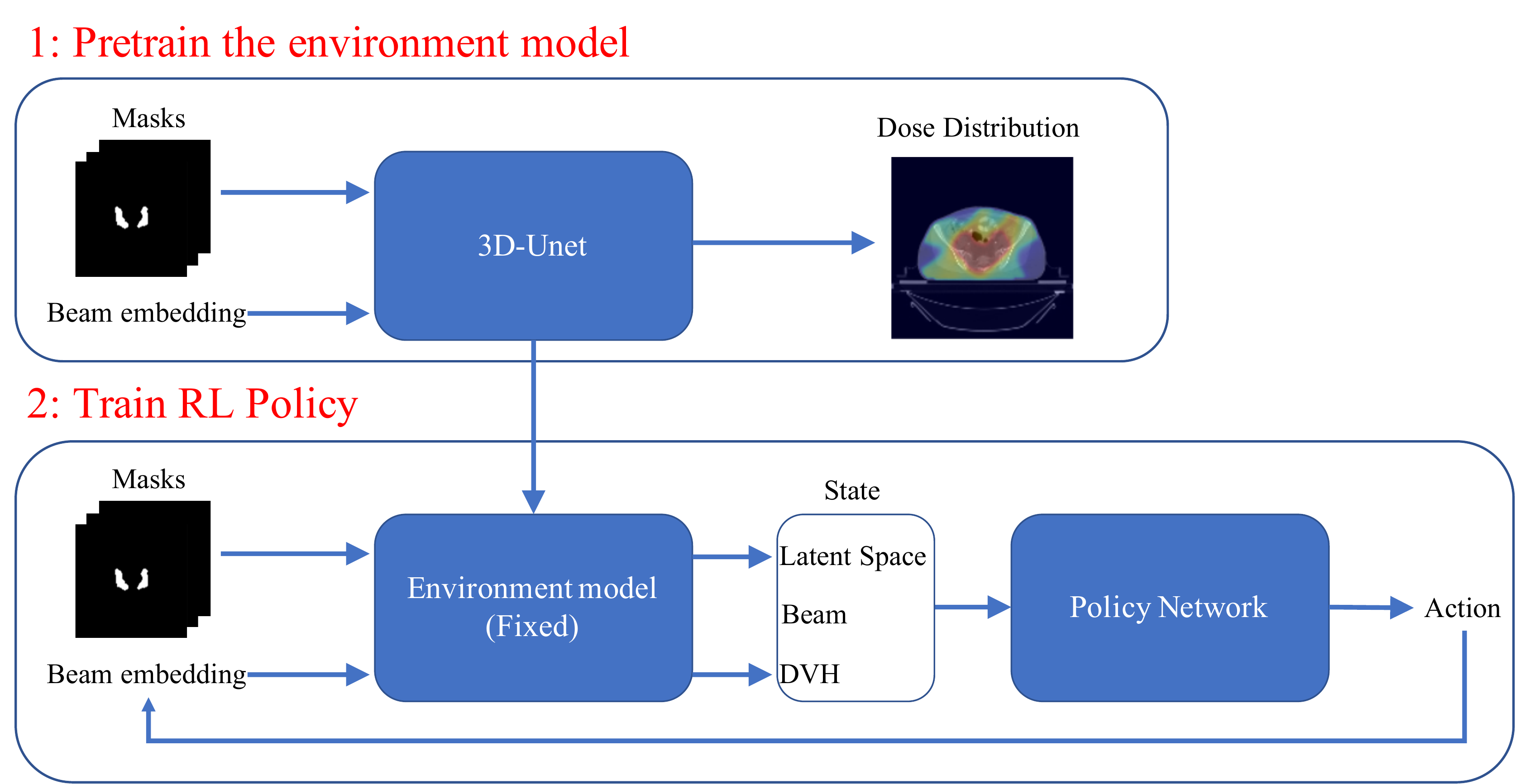}
    \caption{The overall pipeline of the proposed method.}
    \label{fig:pipeline}
\end{figure*}

To improve the quality of IMRT treatment planning, we propose a DRL-based approach to BAO, where the task is regarded as a sequential decision-making problem and is formulated as an MDP. To facilitate the training process and obtain the next state, we design a 3D-Unet to simulate the IMRT environment. The overall pipeline of our approach is illustrated in Figure \ref{fig:pipeline}. By leveraging the simulation model, we train agents using DDQN and PPO to select the beam angle sequentially, given the patient's anatomic structure and the current treatment plan.

\subsection{MDP Formulation of IMRT BAO}
\label{subsec:MDP}
IMRT BAO problem is a combinatorial optimization problem that can be formulated as a sequential decision-making problem and effectively solved using DRL. In this regard, we need to select one beam angle based on the current state at each time step. Given the patient's segmentation masks $m$, the total number of all possible angles $N$, the MDP formulation of the BAO problem consists of the following components:
\begin{itemize}
    \item The \textbf{state} is a tuple $s_t=(\theta_t, dvh_t, l_t)$. 
    $\theta_t$ records the beam angles that are already selected at time step $t$. It is a binary vector with the size of $N$. The value at each entry equals 1 if the corresponding beam angle is selected. In this paper, we set $N$ to 180, which indicates the total number of candidate beam angles at 2-degree resolution.
    $dvh_t$ denotes the DVH of the current treatment plan based on the selected beam angles. $l_t$ encodes the latent features of the patient's segmentation mask $m$ and the selected beam angles $\theta_{t-1}$.
    
    \item The \textbf{action} $a_t$ is a one-hot vector with the size of $N$, which represents the beam angle that we choose at time step $t$.
    
    \item The \textbf{reward} is calculated as $r_t=CI_t - CI_{t-1}$, where $CI$ denotes the conformity index that can evaluate the quality of the current treatment plan to a certain degree \cite{van1997conformation}. Specifically, $CI$ can be calculated as follows:
        \begin{equation}
            CI = \frac{V_{PTV,ref}^2}{V_{PTV}V_{ref}},
        \end{equation}
    where $V_{PTV}$ denotes the number of voxels of PTV, $V_{ref}$ denotes the number of voxels that received reference dose, and $V_{PTV, ref}$ is the number of voxels of PTV that received reference dose. $\frac{V_{PTV,ref}}{V_{PTV}}$ and $\frac{V_{PTV,ref}}{V_{ref}}$ can describe the PTV coverage rate and the OAR sparing rate respectively. $CI\in[0,1]$, the closer $CI$ is to 1, the more conformal the treatment plan is.
    
    \item The \textbf{transition model} $\mathbb{P}$ is detailed as follows: given the state $s_t$ and $a_t$, each component of next state $s_{t+1}$ can be calculated as follows:
        \begin{itemize}
            \item The new beam angles set $\theta_{t+1}=\theta_t+a_t$. The MDP terminates once the number of selected beam angles reaches max number $B$. 
            \item The new DVH $dvh_{t+1}$ is calculated according to the dose distribution of the current treatment plan. However, to obtain the dose distribution, the FMO problem needs to be solved first, which is time-consuming. Therefore, we design a 3D-Unet to predict the dose distribution for different beam settings, which is detailed in Section \ref{subsec:3DUnet}.
            \item The new feature $l_{t+1}$ is the latent space of the 3D-Unet that tasks the patient's segmentation mask $m$ and beam angle set $\theta_t$ as inputs.
        \end{itemize}

\end{itemize}

\begin{figure*}[t]
    \centering
    \subfloat[Beam extractor.\label{fig:Beam_extractor}]{
    \includegraphics[height=1.5in]{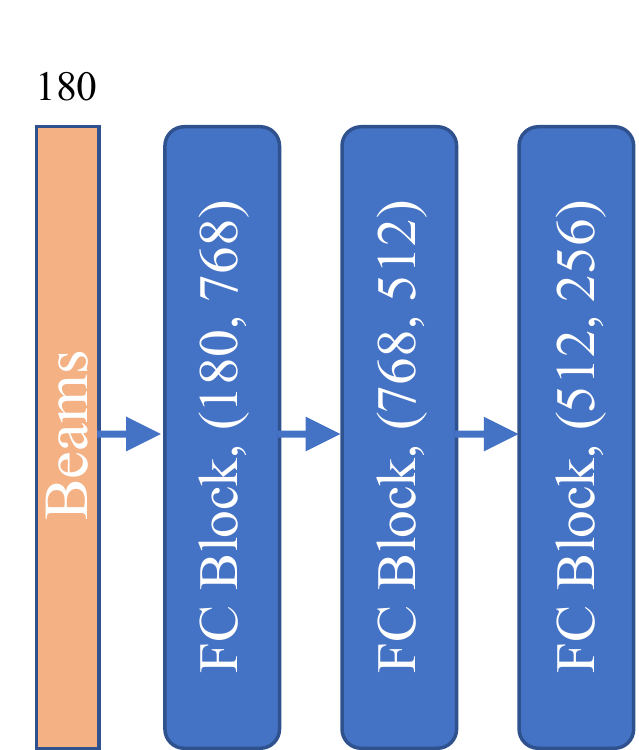}
    }
    \quad\quad\quad
    \subfloat[DVH extractor.\label{fig:DVH_extractor}]{
    \includegraphics[height=1.5in]{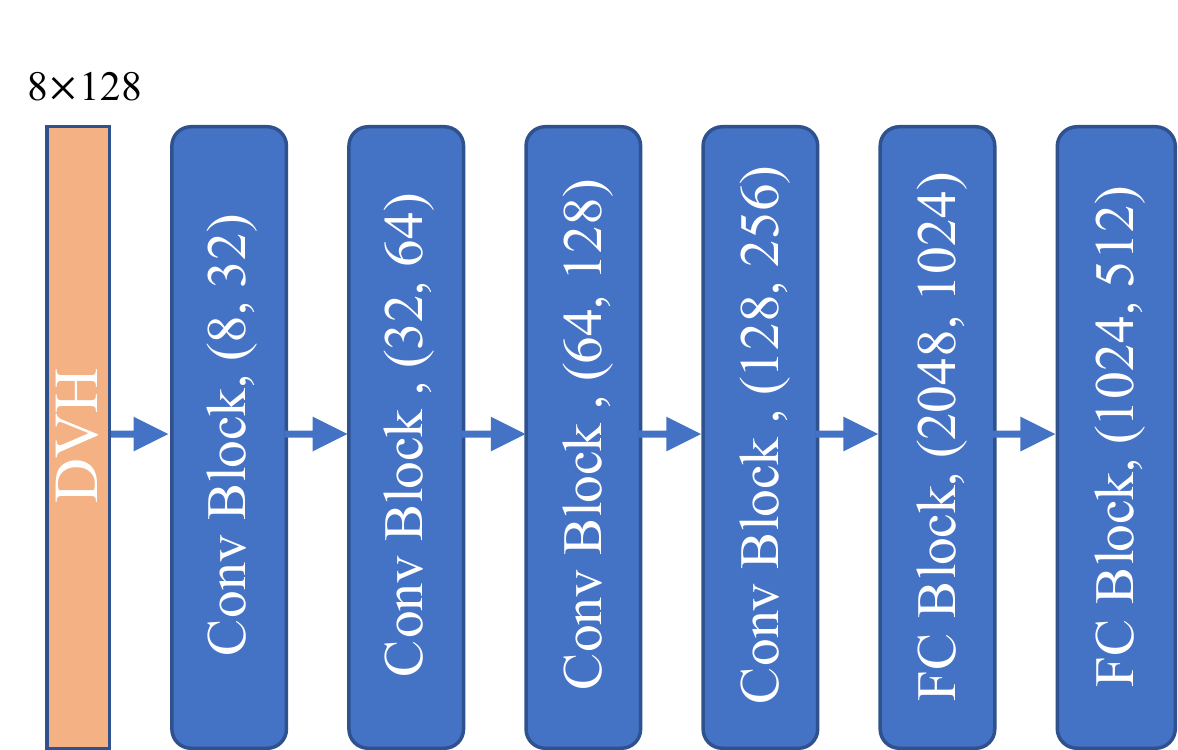}
    }
    \quad\quad\quad
    \subfloat[Latent extractor.\label{fig:Latent_extractor}]{
    \includegraphics[height=1.5in]{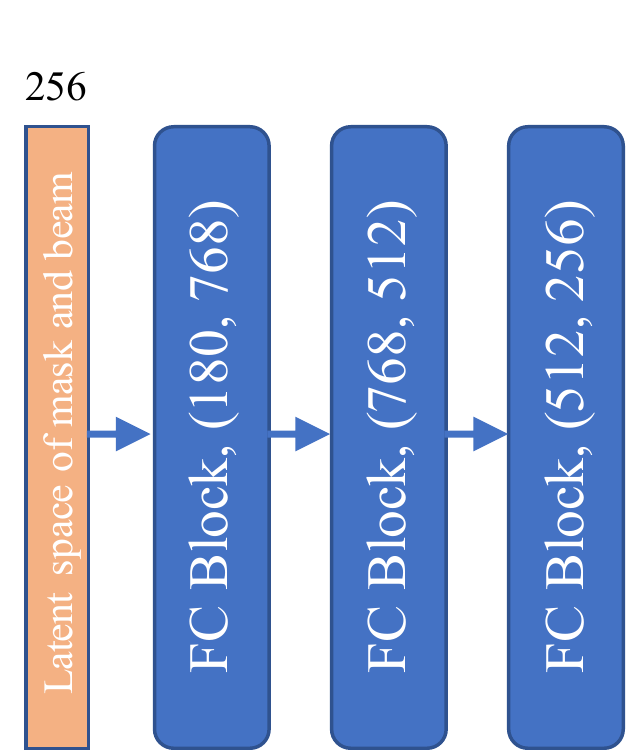}
    }
    \\
    \subfloat[Q-Network.\label{fig:Qnetwork}]{
    \includegraphics[height=1.5in]{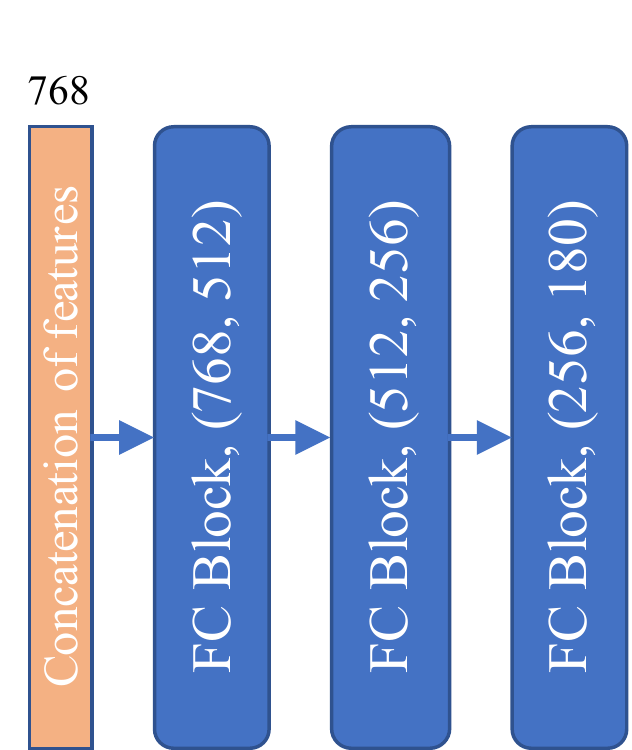}
    }
    \quad
    \subfloat[PPO network.\label{fig:PPONetwork}]{
    \includegraphics[height=2in]{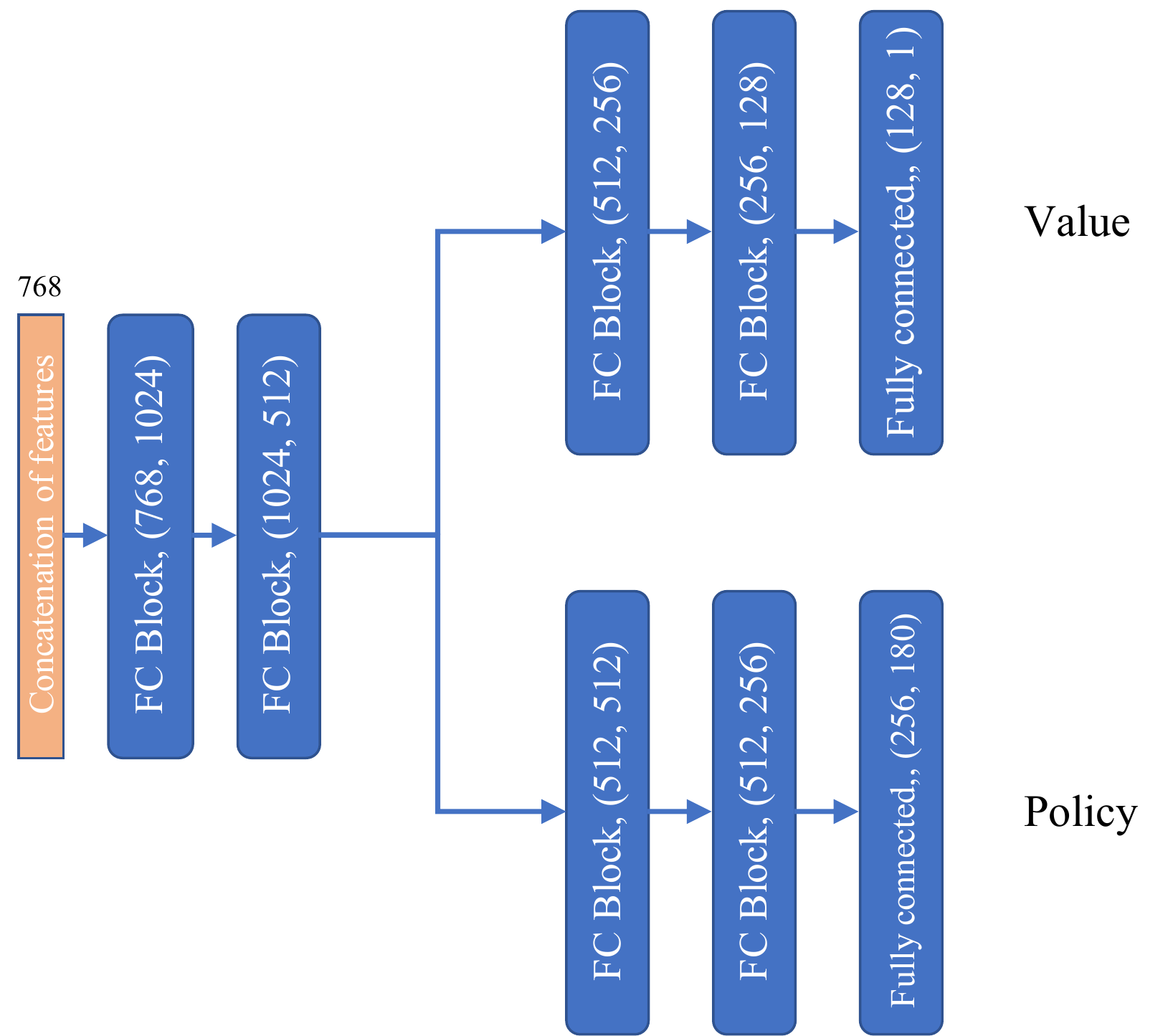}
    }
    \subfloat[Convolutional block.\label{fig:ConvBlock}]{
    \includegraphics[height=1.2in]{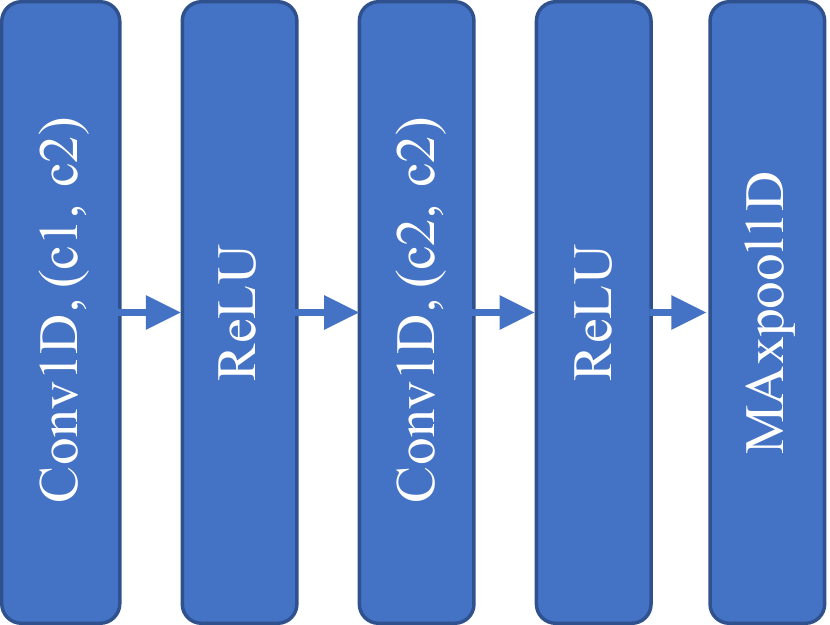}
    }
    \quad\quad
    \subfloat[Fully connected block.\label{fig:FCBlock}]{
    \includegraphics[height=1.2in]{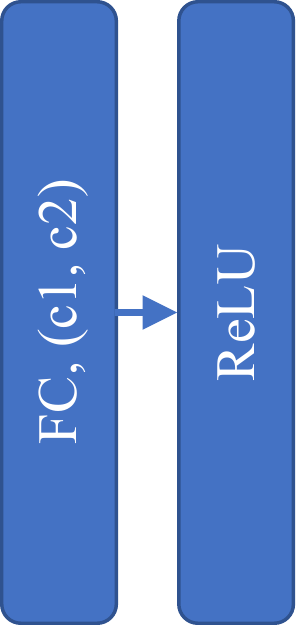}
    }
    \caption{The components of policy networks.}
    \label{fig:policy_network}
\end{figure*}

\subsection{3D-Unet-based IMRT Environment Simulation}
\label{subsec:3DUnet}

As mentioned in Section \ref{subsec:MDP}, to obtain the next state, we need to design a network that can predict the dose distribution for the different number of beam angles. In \cite{barragan2019three}, the authors propose to use ray-tracing dose distribution, which is time-consuming to generate, as the input of the network to represent the beam setup information. In contrast, we propose a 3D-Unet \cite{ronneberger2015u} that directly takes the beam set $\theta$ and segmentation masks $m$ as inputs, and outputs the predicted dose distribution. 

As we use cervical cancer as an example in this paper, thus the segmentation mask contains 1 PTV channel, 6 OAR channels, and 1 body channel. It is worth noting that we use the positional encoding technique \cite{vaswani2017attention} to encode the beam angle, allowing the network to fully consider the beam setting. Additionally, we set the max number of beam angles to be selected as 9, which means we have 9 channels for beam information. We then encode each selected beam angle as a tensor with the size of $128\times128\times64$. Therefore, the input of the 3D-Unet is a 4D tensor with the size of $17\times128\times128\times64$, consisting of 8 segmentation mask channels and 9 beam channels, for a total of 17 channels. The output of the 3D-Unet is the predicted dose distribution with the size of $128\times128\times64$. Based on the above settings, the proposed model can predict dose distribution for beam numbers ranging from 1 to 9. The network architecture used in this paper is similar to the one used in \cite{qilin2022feasibility}, but with two key differences. Firstly, we add the SEblocks \cite{hu2018squeeze} before the max-pooling and the transpose-convolutional layers. Secondly, we use instance normalization \cite{ulyanov2016instance} instead of batch normalization since we have a small batch size. We train our 3D-Unet using the mean squared error loss function.

\subsection{Policy Networks}
As described in Section \ref{subsec:MDP}, our proposed MDP for the IMRT BAO problem involves three components in the state, including the selected beam vector, the current DVH, and the latent feature of the masks and beam encoding. To extract the features from these components, we design a feature extractor with three separate branches for the policy network, as illustrated in Figure \ref{fig:policy_network}. Specifically, for the selected beam vector $\theta_t$, we use a three-layer multi-layer perceptron (MLP) to extract its features, as illustrated in Figure \ref{fig:Beam_extractor}. Similarly, for the patient's anatomy information $l_t$, we also use a three-layer MLP to extract its features, as shown in Figure \ref{fig:Latent_extractor}. For the current DVH $dvh_t$, we use several 1D convolutional layers and MLPs to extract its features, as described in Figure \ref{fig:DVH_extractor}. The outputs of these three branches are then concatenated to form a single feature vector. Finally, for DDQN, the combined feature is sent to a three-layer MLP to predict the Q function, as illustrated in Figure \ref{fig:Qnetwork}. For PPO, the combined feature is first sent to a shared two-layer MLP and then the output is sent to the two separate three-layer MLPs to predict the value function and the policy, respectively, as illustrated in Figure \ref{fig:PPONetwork}.

\section{Experiments}
To demonstrate the performance of the proposed pipeline for IMRT BAO, we conduct two kinds of experiments, including IMRT environment simulation and IMRT BAO. Sixty-seven cervical cancer patients originally treated with radiotherapy in our institute are included. Fifty cases are randomly selected as the training data, 
and the remaining 17 cases are as the testing data. The anatomy information consists of 8 masks, including PTV, bladder, rectum, left and right femoral heads, colon, small intestine, and body. The prescription dose is 50.4 Gy. The patient mask and the dose distribution are down-sampling to $128\times128\times64$. We use several commonly used metrics in the clinic to evaluate the treatment plan, including CI, the dose to 95\% of the target (D$_{95}$), the volume receiving a dose at least 45 Gy (V$_{45}$).

\label{sec:experiments}
\subsection{IMRT Environment Simulation}
\begin{figure*}[t]
    \centering
    \subfloat[One beam angle.\label{fig:one_beam}]{
    \includegraphics[width=2.2in]{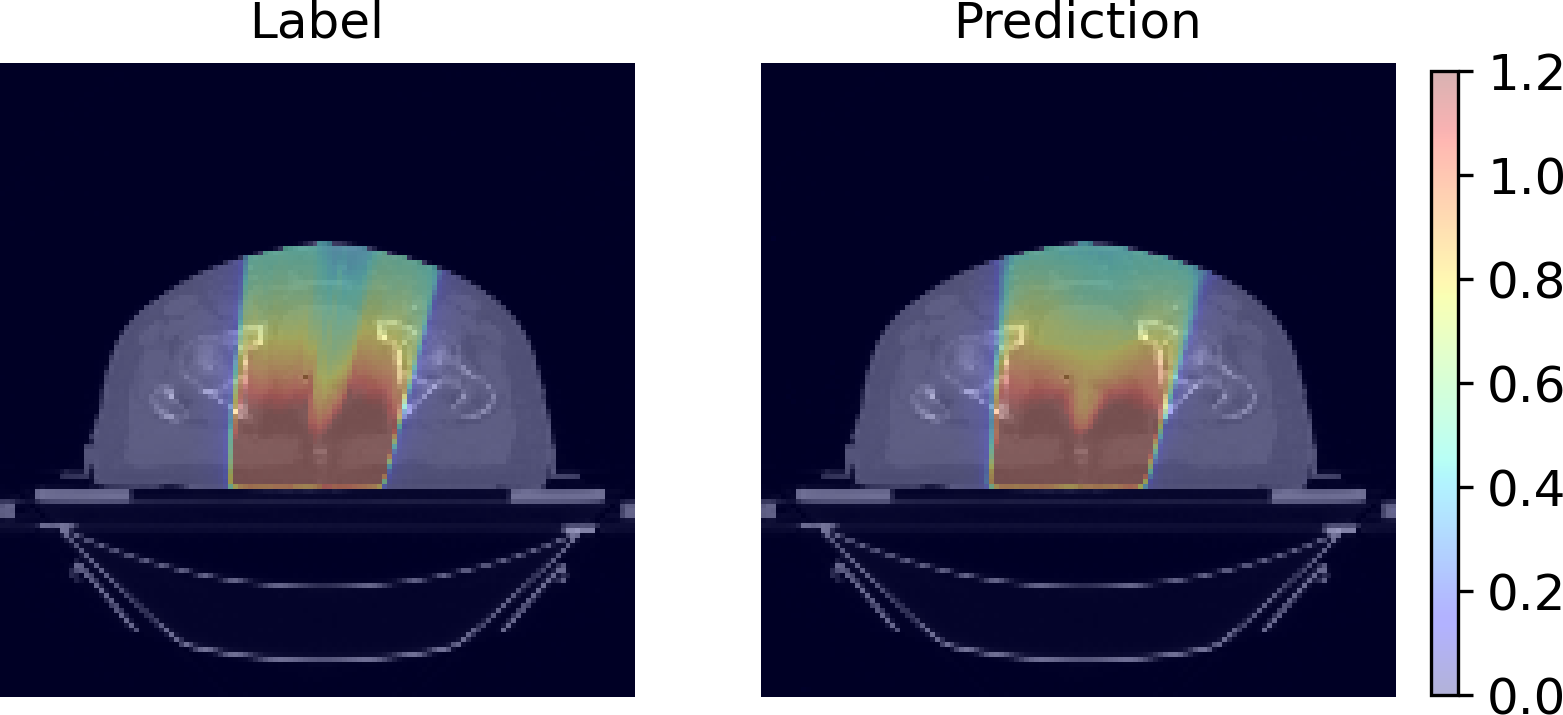}
    }
    \subfloat[Two beam angles.\label{fig:two_beam}]{
    \includegraphics[width=2.2in]{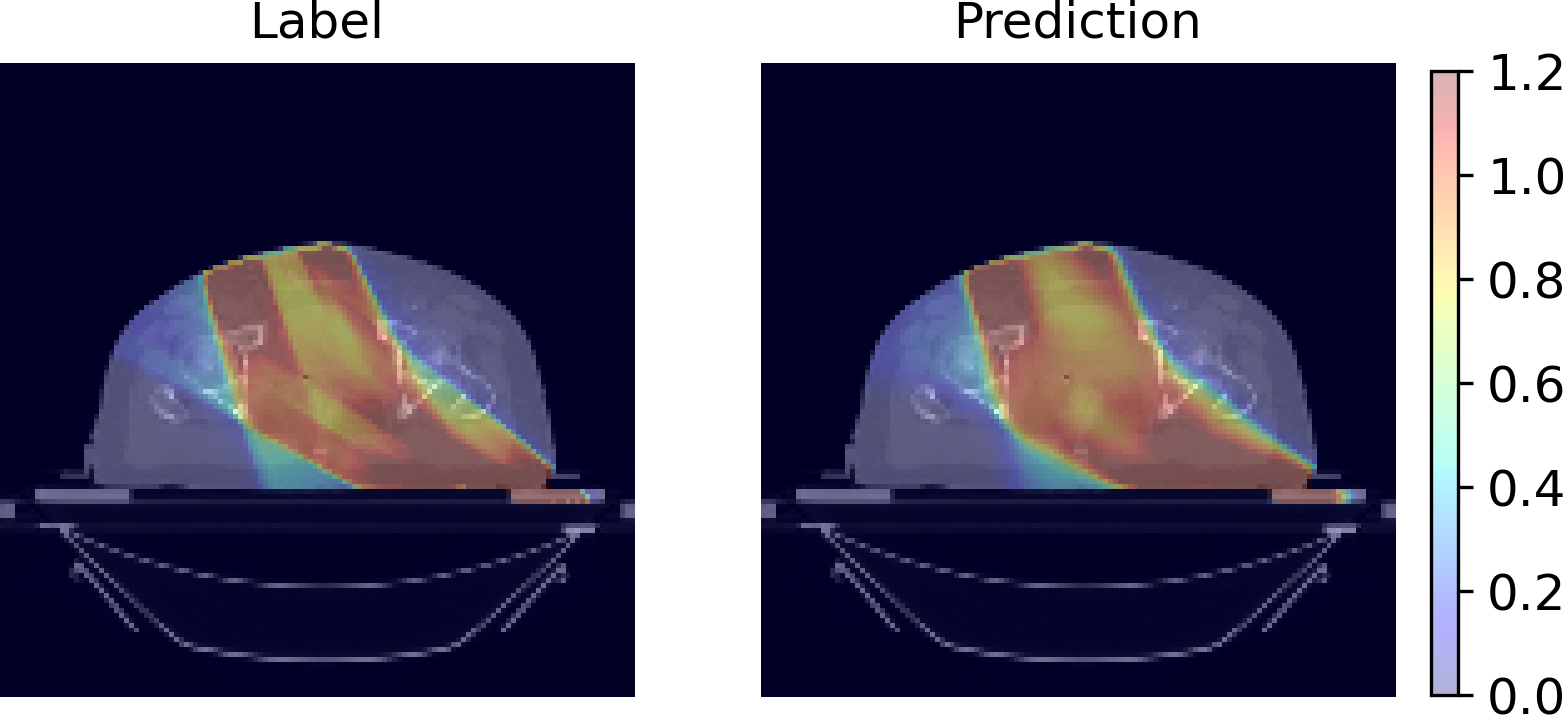}
    }
    \subfloat[Three beam angles.\label{fig:three_beam}]{
    \includegraphics[width=2.2in]{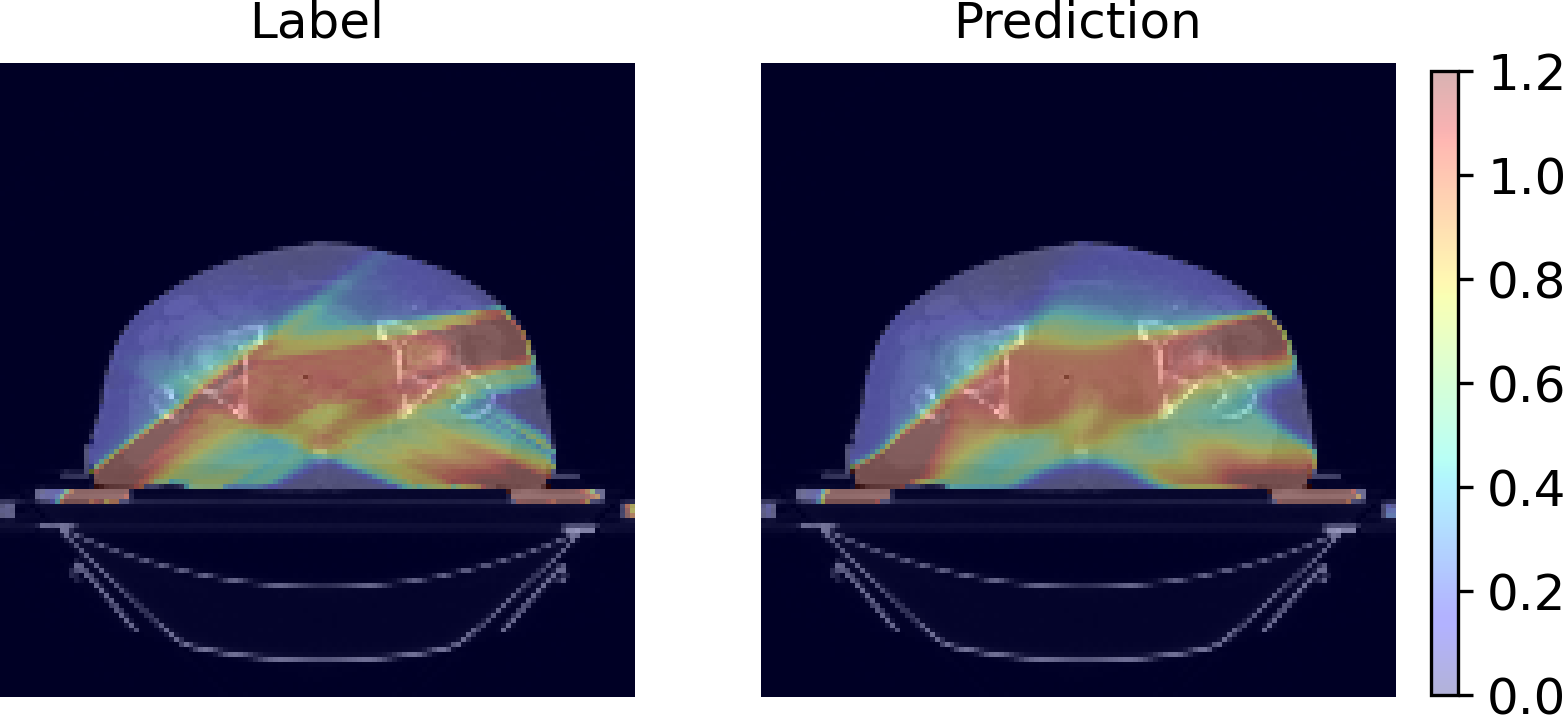}
    }
    \\
    \subfloat[Four beam angles.\label{fig:four_beam}]{
    \includegraphics[width=2.2in]{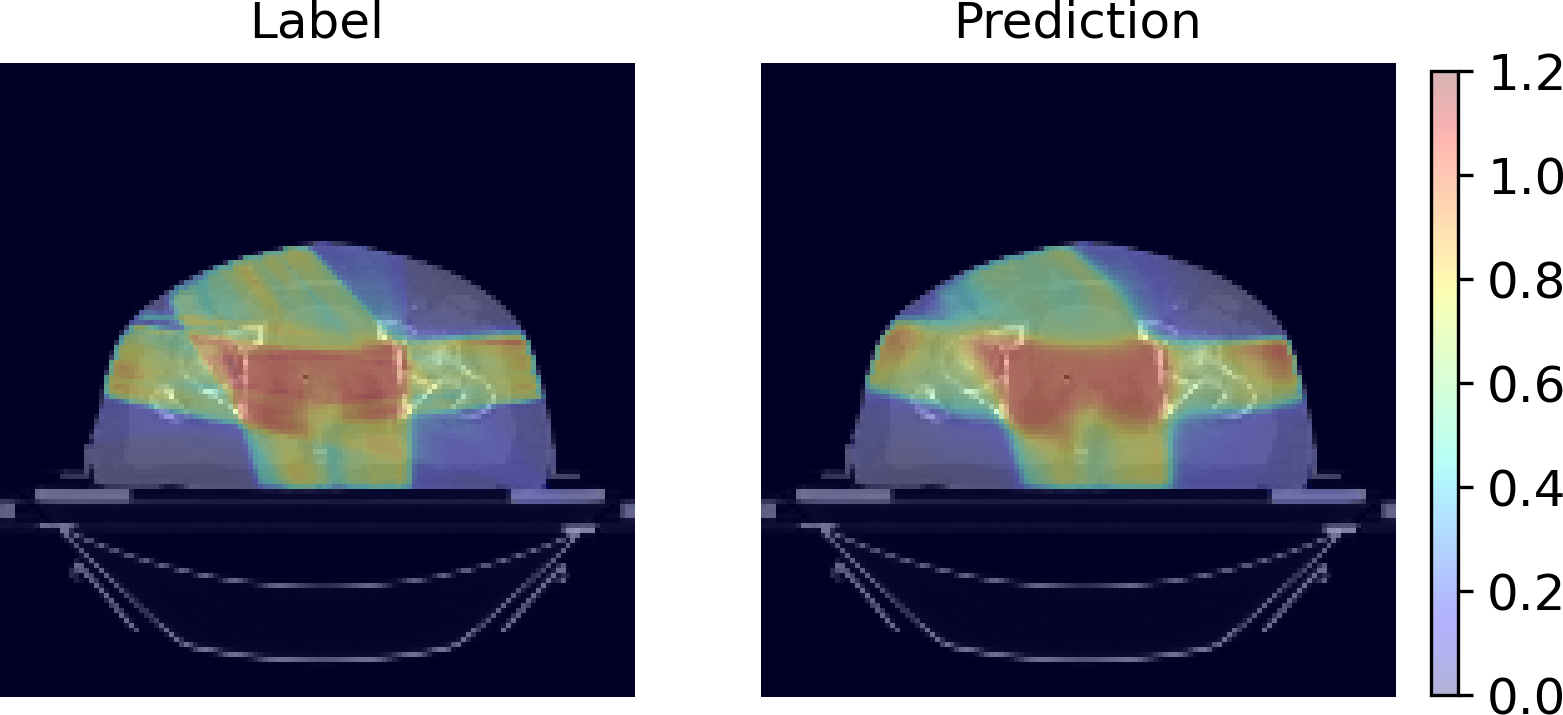}
    }
    \subfloat[Five beam angles.\label{fig:five_beam}]{
    \includegraphics[width=2.2in]{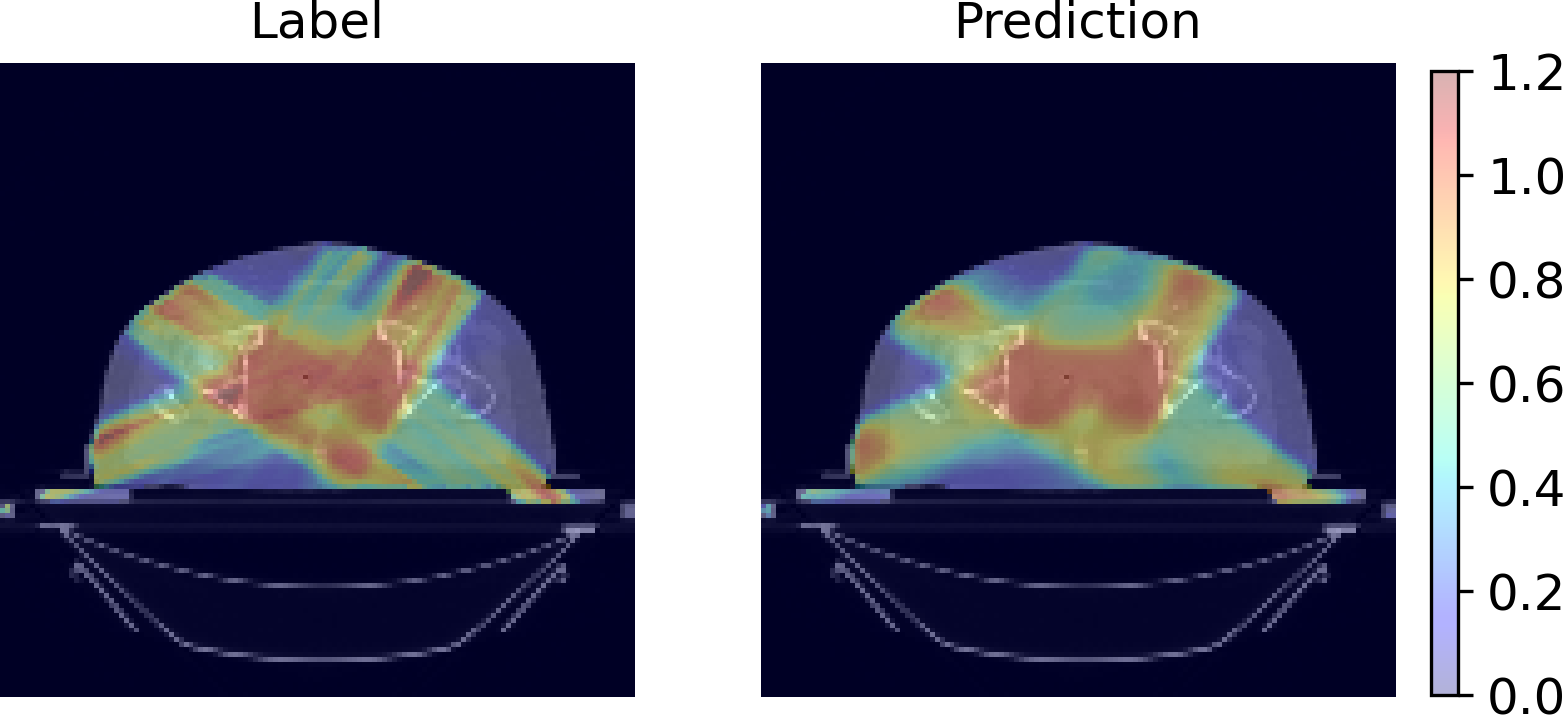}
    }
    \subfloat[Six beam angles.\label{fig:six_beam}]{
    \includegraphics[width=2.2in]{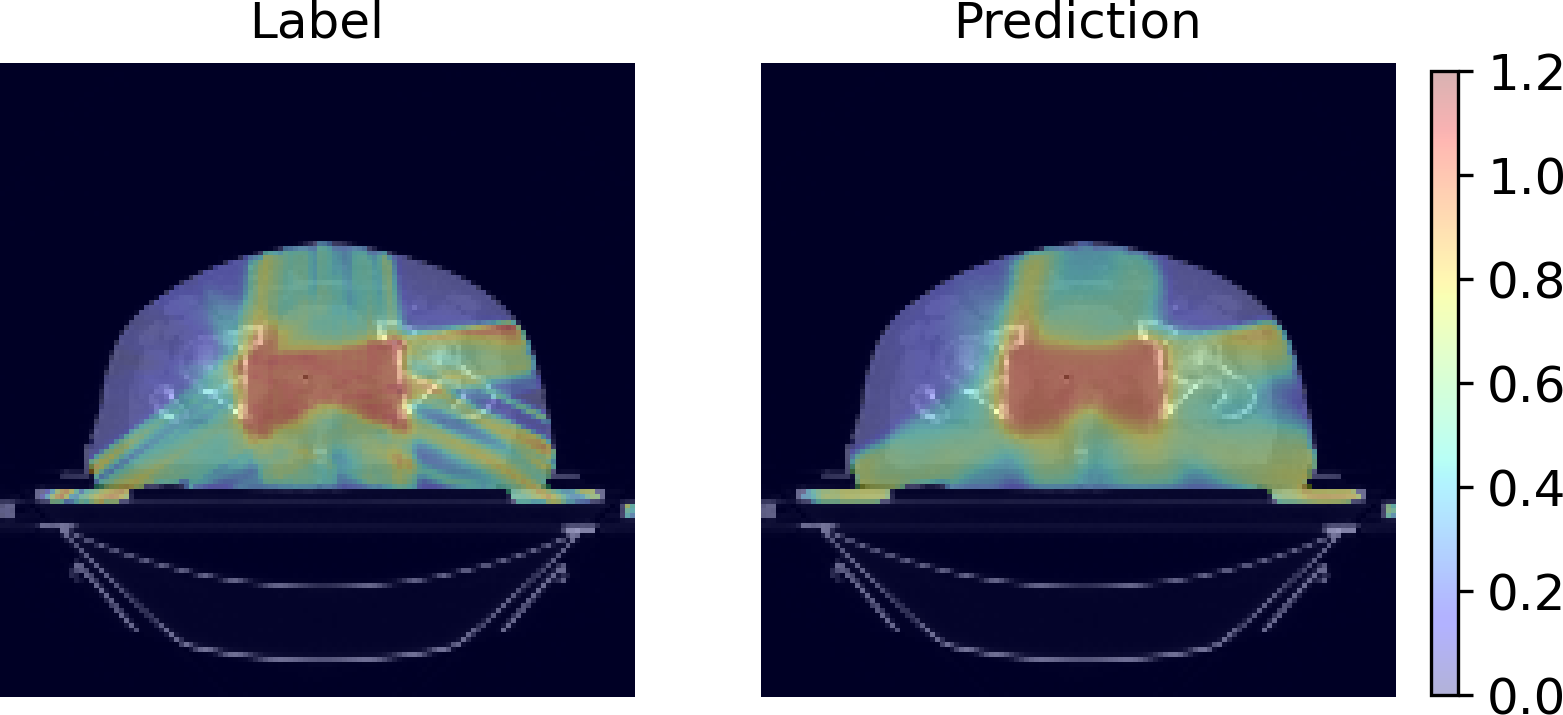}
    }
    \\
    \subfloat[Seven beam angles.\label{fig:seven_beam}]{
    \includegraphics[width=2.2in]{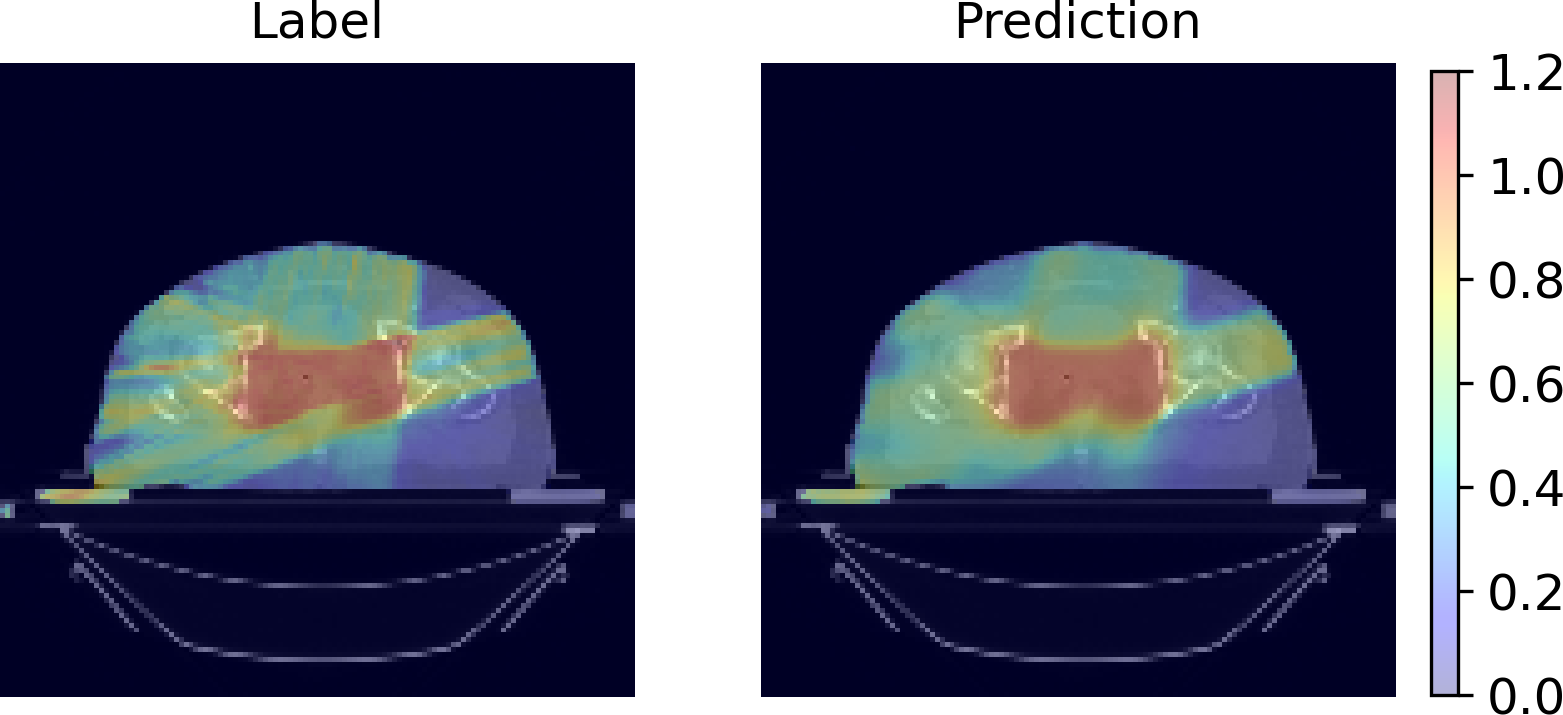}
    }
    \subfloat[Eight beam angles.\label{fig:eight_beam}]{
    \includegraphics[width=2.2in]{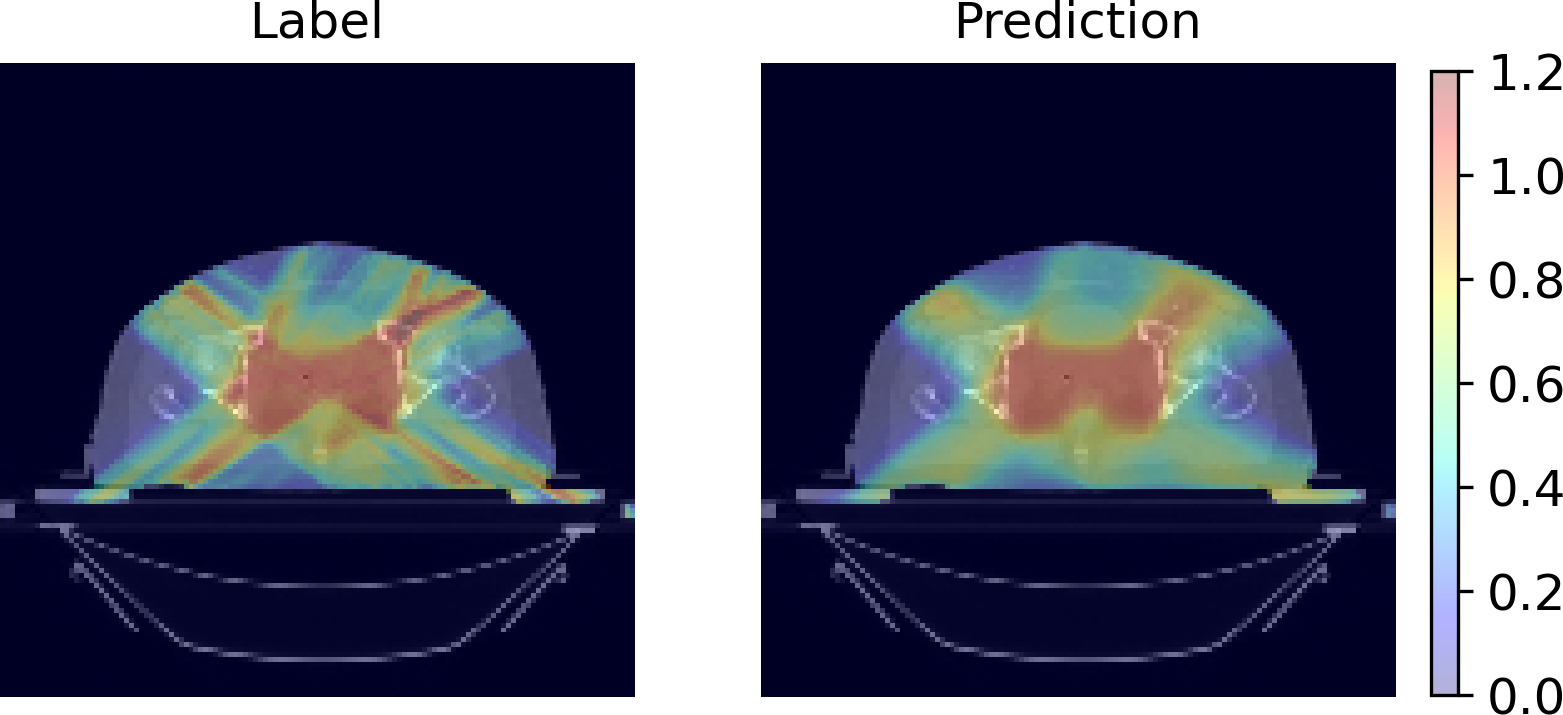}
    }
    \subfloat[Nine beam angles.\label{fig:nine_beam}]{
    \includegraphics[width=2.2in]{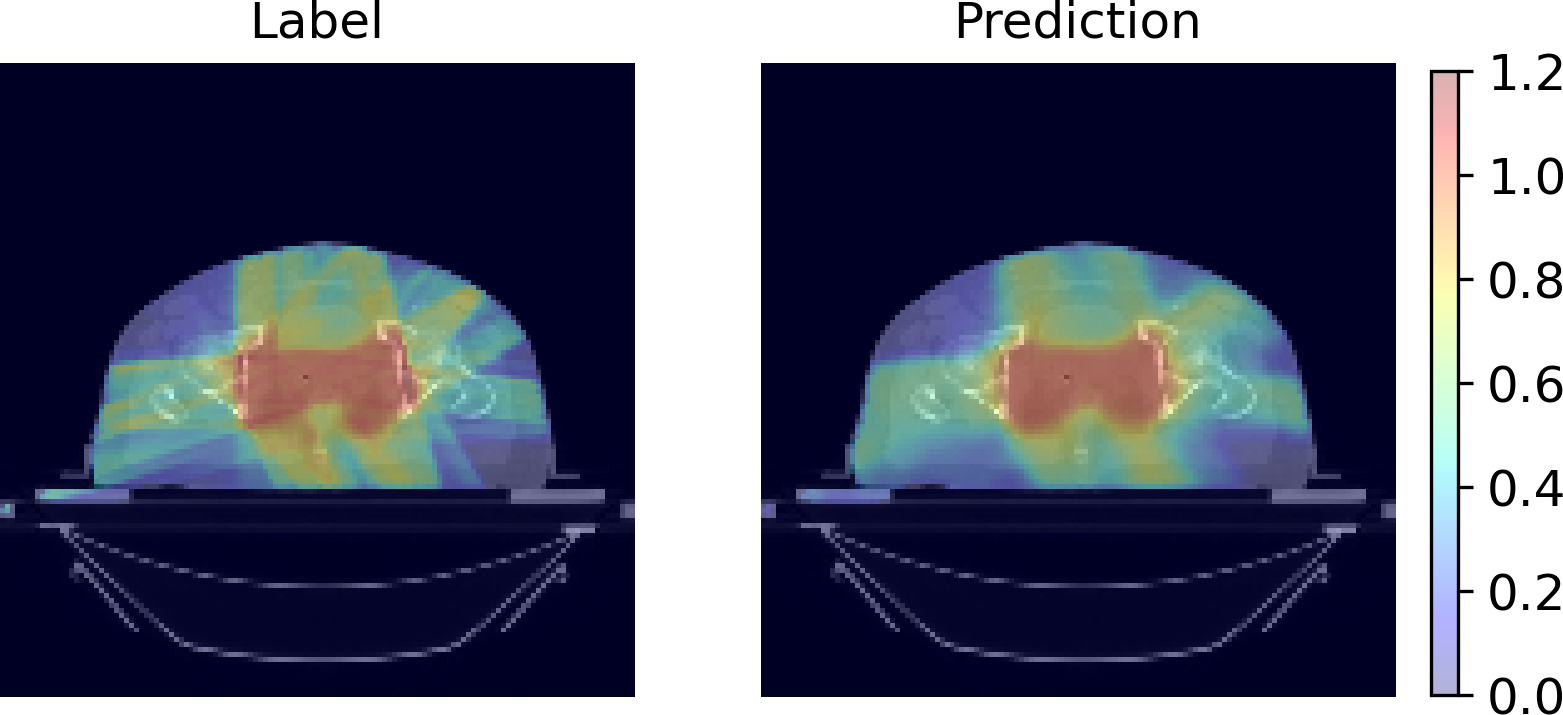}
    }
    \caption{Results of IMRT environment simulation. In each sub-figure, the Label denotes the ground truth and the Prediction denotes the output of the 3D-Unet.}
    \label{fig:dose_prediciton}
\end{figure*}

\begin{table*}[t]
    \centering
    \begin{tabular}{ccccccc}
        \toprule[1.5pt]
        & \multicolumn{6}{c}{Mean absolute error in percentages (\%)} \\
        Num. Angles & PTV D$_{95}$ & CI & Rectum V$_{45}$ & Bladder V$_{45}$ & Colon V$_{45}$ & Small Intestine V$_{45}$\\
        \hline
        1 & 4.41$\pm$2.66 & 0.71$\pm$0.56 & 7.56$\pm$8.78 & 5.45$\pm$6.81 & 1.20$\pm$1.15 & 0.88$\pm$1.10 \\
        2 & 2.94$\pm$2.27 & 2.05$\pm$1.67 & 9.38$\pm$7.41 & 9.25$\pm$8.32 & 1.94$\pm$2.14 & 2.50$\pm$2.80 \\
        3 & 1.59$\pm$1.38 & 2.71$\pm$2.37 & 6.67$\pm$5.23 & 6.13$\pm$5.96 & 2.62$\pm$2.66 & 2.97$\pm$2.74 \\
        4 & 0.85$\pm$0.91 & 4.65$\pm$3.62 & 5.05$\pm$4.04 & 4.27$\pm$4.24 & 2.28$\pm$2.76 & 3.36$\pm$3.60 \\
        5 & 0.61$\pm$0.58 & 5.09$\pm$3.66 & 4.51$\pm$3.78 & 3.49$\pm$3.21 & 2.10$\pm$2.81 & 2.98$\pm$3.56 \\
        6 & 0.47$\pm$0.37 & 5.74$\pm$3.83 & 4.17$\pm$3.41 & 3.11$\pm$2.75 & 1.74$\pm$2.08 & 2.42$\pm$2.95 \\
        7 & 0.49$\pm$0.37 & 6.30$\pm$5.97 & 3.96$\pm$2.98 & 3.06$\pm$2.53 & 1.75$\pm$3.05 & 3.04$\pm$5.46 \\
        8 & 0.54$\pm$0.42 & 6.00$\pm$6.28 & 3.56$\pm$2.73 & 2.81$\pm$2.71 & 1.70$\pm$3.36 & 2.90$\pm$5.32 \\
        9 & 0.53$\pm$0.41 & 5.44$\pm$6.64 & 3.43$\pm$2.81 & 2.78$\pm$2.54 & 1.74$\pm$3.37 & 2.67$\pm$5.48 \\
        Overall & 1.38$\pm$1.87 & 4.30$\pm$4.72 & 5.37$\pm$5.37 & 4.48$\pm$5.20 & 1.90$\pm$2.71 & 2.64$\pm$3.99 \\
        \toprule[1.5pt]
    \end{tabular}
    \caption{The mean absolute error and its standard deviation (mean$\pm$std) for clinically relevant metrics on the testing set (average results for 5 folds). All results are expressed in percentages. Both overall results and the results of the different number of beam angles are reported.}
    \label{tab:dose_prediction}
\end{table*}

\subsubsection{Experiment Setup}
In this section, we conduct an experiment to demonstrate that the proposed IMRT simulation model can predict dose distribution for different beam settings. Specifically, we use an open-source toolkit MatRad \cite{wieser2017development} to generate datasets for training, validation, and testing. For each case, we randomly generate 80 plans for each number of the beam setting, ranging from 1-9, which means 720 (80$\times$9) plans per case, resulting in a total of 48240 (720$\times$67) plans. The possible angles are all integers in $[0^\circ, 360^\circ)$ with the resolution of $2^\circ$, resulting in 180 possible angles. Five-fold cross-validation is used to evaluate the model performance. The proposed 3D-Unet is trained with six graphics processing units NVIDIA TITAN V.
And We use Adam \cite{kingma2014adam} to optimize the parameters of the 3D-Unet with an initial learning rate of 0.0001. The learning rate is scheduled according to the cosine annealing method. After training, we test the learned prediction models on 12240 plans of 17 testing cases. 

\subsubsection{Results}
In Figure \ref{fig:dose_prediciton}, we visualize some prediction results. We can find that the prediction results are similar to the label data, which demonstrates that the proposed 3D-Unet can capture the beam angle information for the model input and predict the dose distribution for different beam settings. Furthermore, Table \ref{tab:dose_prediction} reports the mean absolute error (MAE) of some commonly used metrics in the clinic to evaluate the model prediction. The overall MAEs are less than 5\% for most metrics, except Rectum, which is slightly bigger than 5\%. In addition, we can notice that the error is relatively large only when the number of angles is small. With this prediction model, we can predict the dose distribution for the different number of beam angles less than 1s. Besides, we can make predictions in parallel which significantly reduces the time to train the RL agent.

\subsection{IMRT Beam Angle Optimization}

\begin{table}[t]
    \centering
    \begin{tabular}{cccc}
        \toprule[1.5pt]
         & \multicolumn{3}{c}{Num. Angles} \\
        \hline
        Method & 5 & 7 & 9 \\
        \hline
        DDQN & 0.7702$\pm$0.0225 & 0.7956$\pm$0.0269 & \textbf{0.8221$\pm$0.0218} \\
        PPO & \textbf{0.7829$\pm$0.0410} & \textbf{0.8039$\pm$0.0353} & 0.8191$\pm$0.0373 \\
        Even & 0.7449$\pm$0.0259 & 0.7780$\pm$0.0247 & 0.7979$\pm$0.0239 \\
        \toprule[1.5pt]
    \end{tabular}
    \caption{The mean CI and its standard deviation (mean$\pm$std). The results are computed based on the dose distribution predicted by the environment model the prescription dose is 50.4 Gy. The best results for different beam numbers are shown in bold numbers. }
    \label{tab:CI_pred}
\end{table}

\begin{table*}[t]
    \centering
    \begin{tabular}{ccccccc}
        \toprule[1.5pt]
        & \multicolumn{3}{c}{CI} & \multicolumn{3}{c}{PTV D$_{95}$} \\
        \hline
        Method & 5 & 7 & 9 & 5 & 7 & 9 \\
        \hline
        DDQN & 0.6754$\pm$0.0431 & 0.7138$\pm$0.0586 & \textbf{0.7520$\pm$0.0465} & 50.0927$\pm$0.5532 & 50.5369$\pm$ 0.3381 & 50.6762$\pm$0.2249 \\
        PPO & \textbf{0.6932$\pm$0.0383} & \textbf{0.7337$\pm$0.0459} & 0.7479$\pm$0.0740 & 50.1135$\pm$0.4577 & \textbf{50.4772$\pm$0.3391} & \textbf{50.6318$\pm$0.3045} \\
        Even & 0.6482$\pm$0.0705 & 0.6823$\pm$0.0761 & 0.7220$\pm$0.0691 & \textbf{50.2062$\pm$0.3944} & 50.5985$\pm$0.2947 & 50.7103$\pm$0.2393 \\
        \hline
        & \multicolumn{3}{c}{Rectum V$_{45}$} & \multicolumn{3}{c}{Bladder V$_{45}$} \\
        \hline
        Method & 5 & 7 & 9 & 5 & 7 & 9 \\
        \hline
        DDQN & \textbf{0.3233$\pm$0.0433} & \textbf{0.3273$\pm$0.0427} & \textbf{0.3069 $\pm$0.0536} & 0.3505$\pm$0.0281  & \textbf{0.3442$\pm$0.0331} & 0.3433$\pm$0.0358 \\
        PPO & 0.3299$\pm$0.0414 & 0.3287$\pm$0.0436 & 0.3143$\pm$0.0452 & \textbf{0.3494$\pm$0.0310} & 0.3490$\pm$0.0311 & 0.3457$\pm$0.0372 \\
        Even & 0.3405$\pm$0.0443 & 0.3303$\pm$0.0453 & 0.3117$\pm$0.0492 & 0.3538$\pm$0.0.0228 & \textbf{0.3442$\pm$0.0322} & \textbf{0.3416$\pm$0.0371} \\
        \hline
        & \multicolumn{3}{c}{Colon V$_{45}$} & \multicolumn{3}{c}{Small Intestine V$_{45}$} \\
        \hline
        Method & 5 & 7 & 9 & 5 & 7 & 9 \\
        \hline
        DDQN & \textbf{0.3019$\pm$0.1545} & \textbf{0.2912$\pm$0.1569} & 0.2796$\pm$0.1515 & \textbf{0.1904$\pm$0.1081} & 0.1885$\pm$0.1170 & 0.1702$\pm$0.0944 \\
        PPO & 0.3086$\pm$0.1593 & 0.2924$\pm$0.1559 & \textbf{0.2778$\pm$0.1532} & 0.1971$\pm$0.0904 & \textbf{0.1721$\pm$0.0803} & \textbf{0.1693$\pm$0.0797} \\
        Even & 0.3095$\pm$0.1619 & 0.3012$\pm$0.1690 & 0.2879$\pm$0.1535 & 0.2139$\pm$0.1061 & 0.2086$\pm$0.1297 & 0.1768$\pm$0.0878 \\
        \toprule[1.5pt]
    \end{tabular}
    \caption{The mean and standard deviation of metrics (mean$\pm$std). The results are computed based on the dose distribution optimized in MatRad and the prescription dose is 50.4 Gy. The best results for different beam numbers are shown in bold numbers. }
    \label{tab:BAO}
\end{table*}

\begin{figure*}[t]
    \centering
    \includegraphics[width=6.6in]{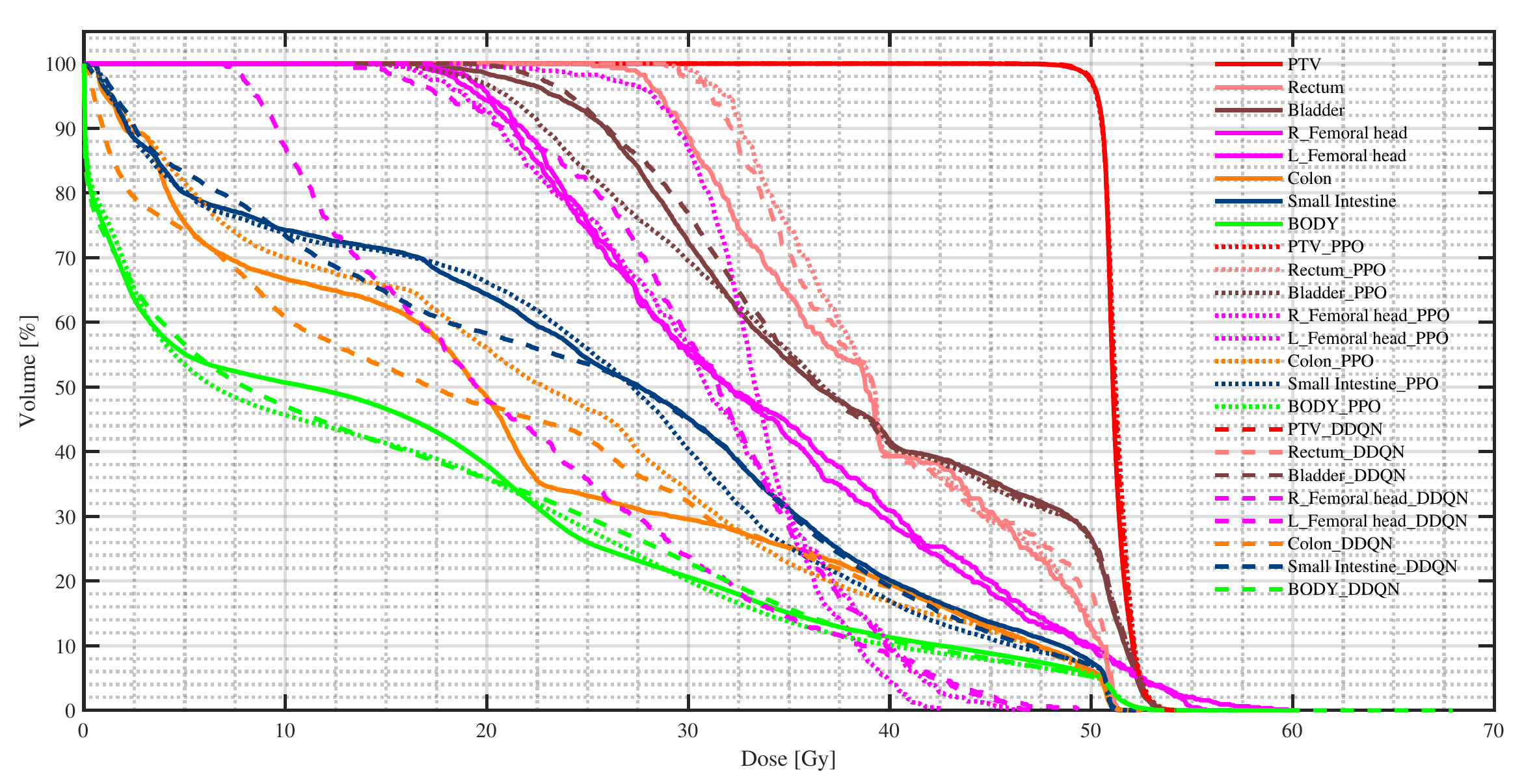}
    \caption{The DVH of a representative case.}
    \label{fig:sample_DVH}
\end{figure*}

\begin{figure*}[t]
    \centering
    \subfloat[DDQN. CI=0.7627. \label{fig:BOO_ddqn}]{
    \includegraphics[width=2in]{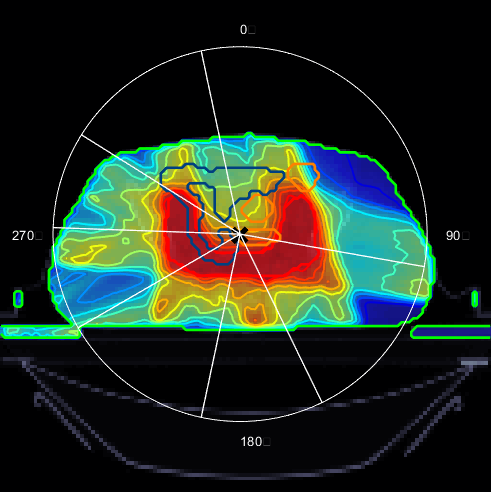}
    }
    \subfloat[PPO. CI=0.7556. \label{fig:BOO_ppo}]{
    \includegraphics[width=2in]{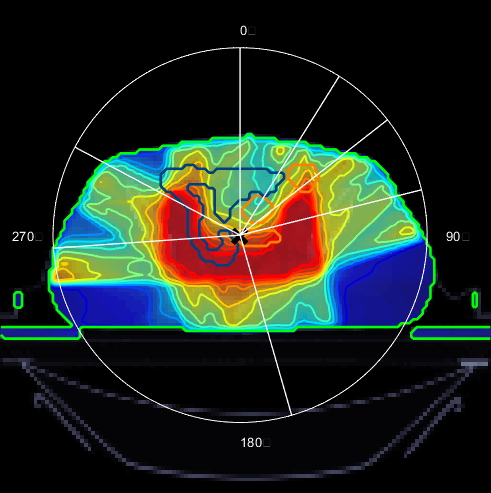}
    }
    \subfloat[Even. CI=0.6728. \label{fig:BOO_even}]{
    \includegraphics[width=2in]{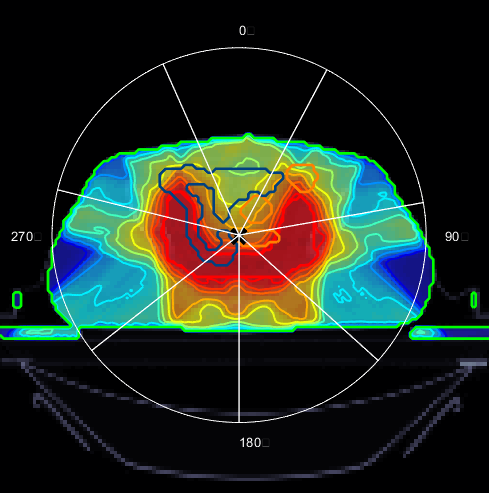}
    }
    \caption{The angle selection of IMRT BAO and its dose distribution in Figure \ref{fig:sample_DVH}.}
    \label{fig:BOO_sample_res}
\end{figure*}

\subsubsection{Experiment Setup}
Based on the pre-trained 3D-Unet as the IMRT environment model, in this section, we use DDQN and PPO to train the agents to determine the beam angles with one graphics processing unit NVIDIA TITAN V. Specifically, we set the number of beam angles to be selected to 5, 7, 9. 
We test the learned policies on 17 testing cases and compare the dose metrics of DRL-based results with those of the evenly distributed beam angles, including CI, PTV D$_{95}$, Rectum V$_{45}$, Bladder V$_{45}$, Colon V$_{45}$ and Small Intestine V$_{45}$. The evenly distributed beam angles start from $180^\circ$, resulting in  [$36^\circ$, $108^\circ$, $180^\circ$, $252^\circ$, $324^\circ$], [$28^\circ$, $80^\circ$, $132^\circ$, $180^\circ$, $232^\circ$, $284^\circ$, $336^\circ$], [$20^\circ$, $60^\circ$, $100^\circ$, $140^\circ$, $180^\circ$, $220^\circ$, $260^\circ$, $300^\circ$, $340^\circ$] for 5, 7, and 9 beam angles respectively.

\subsubsection{Results}
As we use the CI as the reward and train agents based on the simulated environment, we first present the overall mean and standard deviation of CI computed based on the dose distribution predicted by the 3D-Unet in Table \ref{tab:CI_pred}. We can notice that both DDQN and PPO results have improvement on CI for three different beam number settings compared with results of evenly distributed beam angles. To further demonstrate the feasibility of the proposed pipeline, we present the clinically relevant metrics based on the dose distribution optimized by MatRad, which we consider as the real environment in this paper, in Table \ref{tab:BAO}. The results demonstrate that the treatment plans with beam angles selected by DRL perform well in the real environment with regard to CI, and the other important metrics are also comparable. This shows the potential of the proposed DRL-based beam angle selection method in improving the quality of IMRT treatment planning for clinical use.

To better visualize the performance of the proposed pipeline, we present the DVH of a representative case with 7 beam angles in Figure \ref{fig:sample_DVH}. We can notice that DRL-based plans and the evenly distributed-based plan have similar PTV coverage. However, DRL-based plans have smaller high-dose regions for most OARs than the evenly distributed-based plan. This suggests that the DRL-based beam angle selection method can generate plans with improved dose sparing for OARs while maintaining PTV coverage, which is an important factor in reducing treatment-related side effects and improving treatment outcomes. We further present the selected angles and corresponding dose distribution of the case shown in Figure \ref{fig:BOO_sample_res}.

\section{Conclusion}
\label{sec:conclusion}
IMRT BAO is a combinatorial optimization problem and is NP-hard which is difficult to solve. In recent years, the DRL has been proven useful in solving many combinatorial optimization problems, such as TSP and VRP. In this paper, we proposed to use DRL algorithms to learn a personalized beam angle selection strategy for IMRT based on the current treatment plan and the patient's anatomy information. Specifically, we formulated the IMRT BAO problem as an MDP and solved it using DDQN and PPO. To obtain the next state and accelerate the training process, we also designed a 3D-Unet to simulate the IMRT environment. This 3D-Unet can predict the dose distribution for the different number of beam angles within 1 s. With the IMRT environment prediction model, the time spent to train the RL agent is significantly reduced. After training, the agent can determine the personalized beam angles within a few seconds. Experimental results on a dataset of cervical cancer patients showed that the proposed DRL approach can improve the plan quality compared to the strategy used in clinical practice. Specifically, the proposed approach can generate plans with improved dose-sparing for OARs while maintaining PTV coverage. Furthermore, although this paper takes cervical cancer as an example, the proposed method can be also used for other sites. In our future work, we will explore more complex neural network models, such as vision transformer \cite{dosovitskiy2021an, liu2021swin}, to further improve the accuracy of the environment model prediction. Furthermore, we may also consider model-based RL algorithms.

\bibliographystyle{IEEEtran}
\bibliography{main}

\end{document}